\documentclass[preprint,showpacs,preprintnumbers,amsmath,amssymb]{revtex4}

\usepackage{graphicx}
\usepackage{dcolumn}
\usepackage{bm}

\newcommand{\be}{\begin{equation}}  
\newcommand{\ee}{\end{equation}}  
\newcommand{\bear}{\begin{eqnarray}}  
\newcommand{\eear}{\end{eqnarray}}  
\newcommand{\ba}{\begin{array}}  
\newcommand{\ea}{\end{array}}

\newskip\humongous \humongous=0pt plus 1000pt minus 1000pt

\newif\ifdtup

\def\oldreffmt#1{\rlap{[#1]} \hbox to 2\parindent{}}

\def\figfmt#1{\rlap{Figure {#1}} \hbox to 1in{}}  
  
\def\ie{\hbox{\it i.e.}{}}	\def\etc{\hbox{\it etc.}{}}  
\def\eg{\hbox{\it e.g.}{}}

\def\tr{\mathop{\rm tr}}  
\def\Tr{\mathop{\rm Tr}}

\def\beq{\begin{equation}}  
\def\eeq{\end{equation}}  
\def\bea{\begin{eqnarray}}  
\def\eea{\end{eqnarray}}  
\def\half{\frac{1}{2}}  
  
\def\bq{\begin{quote}}  
\def\eq{\end{quote}}

\def\half{\frac{1}{2}}       

\relax  

\newdimen\tdim  
\tdim=\unitlength  
\def\bar{\overline}

\begin{document}

\preprint{FERMILAB-Pub-04/345-T}
\preprint{ANL-HEP-PR-04-129}

\title{Dimensional Deconstruction \\
and Wess--Zumino--Witten Terms}

\author{Christopher T. Hill}

{\email{hill@fnal.gov}}

\affiliation{
 {{Fermi National Accelerator Laboratory}}\\
{{\it P.O. Box 500, Batavia, Illinois 60510, USA}}
}%
\author{Cosmas K. Zachos}
{\email{zachos@anl.gov}}
\affiliation{
 {{ High Energy Physics Division, Argonne National Laboratory}}\\
{{\it            Argonne, Illinois 60439-4815, USA}}
}%
\date{\today}

\begin{abstract}
A new technique is developed for the derivation of the Wess-Zumino-Witten
terms of gauged chiral lagrangians. We start in $D=5$ with 
a pure (mesonless)
Yang-Mills theory, which includes relevant gauge field Chern-Simons terms. The
theory is then compactified, and the effective $D=4$ lagrangian is derived
using lattice techniques, or ``deconstruction,'' where pseudoscalar mesons 
arise from the lattice Wilson links. This yields
the WZW term with the correct Witten coefficient by
way of a simple heuristic argument. We discover a novel WZW term
for singlet currents, that yields the full Goldstone-Wilczek current, 
and a  $U(1)$ axial current for the skyrmion, with
the appropriate anomaly structures.
A more detailed analysis is presented of the dimensional
compactification of Yang-Mills in $D=5$ into a gauged
chiral lagrangian in $D=4$, heeding the 
consistency of
the $D=4$ and $D=5$ Bianchi identities. These dictate a 
novel covariant derivative structure in the $D=4$ gauge theory, yielding
a field strength modified by the addition of commutators of chiral currents. 
The Chern-Simons
term of the pure $D=5$ Yang-Mills theory then devolves into the correct form
of the Wess-Zumino-Witten term with an index (the 
analogue of $N_{colors}=3$) of $N=D=5$. The theory also has
a Skyrme term with a fixed coefficient.

\end{abstract}

\maketitle

\section{\bf Introduction}

There is an intriguing parallel between the $D=5$ pure Yang-Mills theory
and the $D=4$ chiral lagrangian theory of mesons. We first summarize 
features of the $D=5$ Yang-Mills theory.

The pure $SU(N)$ Yang-Mills $D=5$ gauge theory supports a 
topological soliton, unique to $D=5$ \cite{deser1}. 
This soliton is simply the instanton, an $SU(2) $ configuration,
lifted to a time slice in $D=5$, associated
with the nontrivial homotopy class  $\Pi_3(SU(2))$, and
it carries a conserved topological current \cite{topo1,topo2}.
The theory actually has two conserved
topological currents, built out of the gauge fields: firstly an 
adjoint representation current
(only for groups possessing $d$-symbols, e.g., SU(3) and higher), and secondly 
a singlet current (present for all groups). The adjoint current controls 
transitions between the various ways in which the instantonic soliton
can be embedded into the gauge group (e.g., a pure ``I-spin'' embedding can
flip to a ``U-spin'' or ``V-spin'' embedding
for $SU(3)$). The singlet current is identically conserved, 
and yields the topological charge of the soliton.

Each of these currents is topological, and cannot be derived by Noetherian variation
of the gauge kinetic term action. The theory 
must therefore be supplemented with an additional 
Chern-Simons term. The Chern-Simons term that
generates the adjoint current is known as
the ``second Chern character'',
(the $D=5$ generalization of the Deser-Jackiw-Schonfeld-Siegel-Templeton
mass term of $D=3$ \cite{schonfeld,deser}; see also \cite{niemi}). 
Under variation 
of the gauge fields, the second Chern character generates
the adjoint current as a source term in the equation
of motion of the gauge field.  
While not manifestly gauge invariant, under small gauge
transformations (those continuously connected
to the identity), the action containing
the second Chern character is invariant.  By contrast, 
for topologically nontrivial transformations,
the action shifts by an additive numerical factor, and
the coefficient of the Chern character is necessarily quantized so 
the path integral is invariant (\ie, with the proper 
coefficient, this shift in the action is then $2\pi N$) \cite{deser,Wu}.

The singlet current has no associated Chern-Simons term
built out of the gauge fields alone. We presently propose to 
introduce a ``dual variable,'' a vector
potential associated with the instantonic soliton. This
allows us to write a new $U(1)$-gauge
invariant topological term which is analogous to the second Chern character
and which generates the singlet current. 

On the other hand, chiral theories of mesons in $D=4$ based on flavor 
$SU(N)_L\times SU(N)_R$ also possess remarkable, and quite similar, 
topological properties.  The theories support 
the skyrmion solution, which is an $SU(2) $ configuration
and a stable topological object 
(whose core is stabilized when the ``Skyrme'' term is added). 
The skyrmion also reflects the nontrivial $\Pi_3(SU(2))$, and 
it carries a conserved
(modulo anomalies) singlet current, the Goldstone-Wilczek current
\cite{goldstone}, which is interpreted as baryon number.
The chiral theory also contains adjoint representation 
topological currents, conserved modulo anomalies. 
These latter currents again exist only for groups with $d$-symbols, and 
govern transitions
of the embeddings of the skyrmion in the diagonal subgroup $SU(N)$. 
A connection between the instantonic soliton of $D=5$ and the skyrmion 
of $D=4$, through compactification, and a matching 
of their $U(1)$ currents was discussed in 
some detail in ref.(\cite{topo2}). In fact, the full form of the 
Goldstone-Wilczek current can be easily inferred from this matching.

The adjoint topological currents in $D=4$ chiral
theories derive from 
the Wess-Zumino-Witten term. 
Remarkably, the WZW term is neither 
manifestly chirally nor gauge invariant; yet it possesses both symmetries 
for small transformations---those
that are continuously connected to the identity. The overall invariance 
of the path integral 
under large topological chiral and gauge transformations leads again to 
quantization conditions on the WZW term coefficient \cite{Witten}.
The singlet Goldstone-Wilczek current has no corresponding WZW term, but
as will be discussed below, and elaborated
in a companion paper, \cite{hill}, 
such a term can be written, provided the $\sigma$ and $\eta'$
mesons are incorporated into the theory. 
This, in turn, leads to a new singlet axial
vector topological current. 

These topologically interesting aspects of $D=4$ chiral lagrangians
have long been known to
follow from the structure of the theory in one higher dimension.  
In the case of a $D=4$, $SU(N)_L\times SU(N)_R$ nonlinear $\sigma$-model, 
described by an
$N\times N$ unitary matrix field, Witten has shown that the WZW term
can be obtained by promoting the global theory
of mesons to $D=5$, where a certain manifestly chirally invariant
Chern-Simons term occurs, built out of the mesons. One 
then compactifies the fifth-dimension 
term, back to $D=4$. This results in the global 
Wess-Zumino term of $D=4$ \cite{Witten}.  
At this stage, by performing
gauge transformations upon the global object, one can 
infer how to introduce the gauge fields to compensate
the local changes in the WZ term.
This leads to the full Wess-Zumino-Witten term
for the gauged chiral lagrangian, which contains the 
full anomaly structure of the theory. 
The WZW term plays another crucial role, that of locking
the parity of the pion to the parity of space \cite{Witten}.
Certain allowed transitions, 
such as $K+\bar{K}\rightarrow 3 \pi$, which would be
absent without the WZW term, now occur
with a topologically quantized amplitude. 

A conceptual drawback of 
this procedure is that local gauge invariance
is induced into a non-gauge invariant object, {\em a posteriori}.  Local
gauge invariance, however, is a more fundamental symmetry than the
chiral symmetry which it breaks, and it is thus preferable to rely on 
a procedure in which local gauge invariance
is present at the outset.   
Then, upon compactification, one would
require that the meson fields appear with their proper kinetic terms 
and gauging.   
Implementing such a procedure for compactification, we would expect that
the second Chern character  of the $D=5$ pure Yang-Mills theory
morphs into the $D=4$ WZW term.
This approach, at least,
may shed some light on the interplay on the interrelationship
of the various symmetries and topology. 

Indeed, there exists in principle such a procedure, 
the latticization of extra dimensions, \cite{wang},
or ``dimensional deconstruction" \cite{harvard}. 
The approach latticizes only the extra dimensions, yielding
the effective kinetic and interaction terms,
while keeping the $D=4$ subspace
in the continuum. This  is related to the earlier
``transverse lattice'' of Bardeen, Pearson and Rabinovici \cite{bardeen}.
For extra dimensional theories, this is a powerful tool, leading to a 
continuum $D=4$ effective description of a theory
that originated as a pure Yang-Mills theory 
in higher dimensions, with emphasis
on maximal manifest gauge invariance. 
In this approach, one derives
the gauge invariant effective lagrangian for a theory in $D=4$
that is defined by compactification of a theory in higher dimensions.
Starting with pure Yang-Mills in $D=5$, one can thus engineer a $D=4$ gauged
chiral lagrangian. The mesons then appear in the compactified
theory, packaged into exponential chiral fields,
which are the Wilson links associated with the latticization.

In the present work, we study the deconstruction of
the $D=5$ Yang-Mills theory, supplemented with the 
second Chern character, and a new singlet auxiliary term.
We begin with a discussion of
the physical basis of orbifold boundary conditions, and
consideration of the topological
aspects of a gauged chiral lagrangian in $D=4$
through the pure Yang-Mills theory in $D=5$ without mesons. 
Presently, we will show how to derive the WZW
term for a gauged chiral 
lagrangian in $D=4$, by 
matching of the vector potentials and the field strengths of
the $D=5$ Yang-Mills theory onto the relevant operators
in the deconstructed effective lagrangian. 
We begin in the next section, after a review of
$D=5$ Yang-Mills and the Chern-Simons terms, with a simple
heuristic discussion that readily yields the WZW 
term on orbifold compactification to $D=4$. 
We will also
anticipate how a new WZW-like term arises from
the singlet auxiliary  term in the parent $D=5$ 
Yang-Mills theory. This new WZW term generates the
singlet Goldstone-Wilczek current, and a new $U(1)$ axial current
for the skyrmion.

We then study the general issue of
topological deformation of $D=5$ Yang-Mills into 
$D=4$ chiral theories in more detail. 
First, we note that there is
a key element we must address that
is missing in a naive deconstruction, and
which is essential to the propagation of topology
from one geometrical dimension to another. This
is the consistency of {\em the Bianchi 
identities}. The ordinary $D=4$ Bianchi 
identities are always automatically 
satisfied by the
specification that the field strength tensor is a commutator 
of covariant derivatives,
since the $D=4$ relations are simply the 
Jacobi identity for antisymmetrized nested
commutators. However, we'll find that 
there is an additional nontrivial Bianchi 
constraint involving the ``lattice hopping derivative.'' 
This is seen to fail for the first plausible definition
of the hopping derivative in $D=5$.

We thus formulate the Bianchi identities on the 
deconstruction lattice, and we find that
we are able to satisfy them 
with a modified definition
of the $D=4$ covariant derivative. The ordinary
derivative is modified by the addition
of  a vector combination of chiral currents
with a special coefficient of $1/2$.  The
formalism automatically
implements the ``magnetic superconductivity,''
or confinement phase on the orbifold branes, $G_{4\mu} =0$.

The Bianchi-consistent modification implies that the effective  
field strength, $G_{\mu\nu}$, is modified by the  addition 
of terms involving the commutators of chiral currents of the mesons.
This term occurs with a fixed coefficient.
The gauge action is therefore modified as well, and there now appear 
in the classical action two Skyrme terms.  The usual Skyrme term
is generated by the current commutator terms in the field strength
tensor and now has a fixed coefficient. 
Moreover, a new Skyrme term that involves the gauge
field, is also present. We thus conjecture
that this modified theory may tighten the link between the 
instantonic soliton in $D=5$ and
the skyrmion in the deconstructed theory in $D=4$. 
With these terms,
there may exist a ``self-dual,'' and even an analytic
skyrmion solution, matching the instantonic soliton 
at large distances.

Given the new Bianchi-consistent action and field 
strength, the pure Yang-Mills 
second Chern character (CS2) term again goes into the 
WZW term. The resulting WZW term is consistent with
Witten's minimal coefficient, but is larger by
a factor of $5$. Thus, we infer that the dimensionality
of the space-time, $D=5$, appears as the
index of the WZW term in the Bianchi-compliant theory, 
where we normally would install
$N_c=3$, the number of colors of QCD.


\section{The D=5 Pure Yang-Mills Theory and Heuristic Derivation of the
$D=4$
Wess-Zumino-Witten Term}

\vskip 0.15in
\noindent{\bf (i) Preliminaries}
\vskip 0.15in

We start with an $SU(N)$ Yang-Mills gauge theory
in $D=5$. The theory relies on vector potentials, $A_A^a(x)$ 
and coordinates $x^A$, where $(A=0,1,2,3,4)$,
and where $x^\mu$ and $(\mu = 0,1,2,3)$ refers to the
usual space-time dimensions. When we say ``fifth component
of a vector, $x^A$'' we mean, of course, $x^4$.

The covariant derivative is 
\beq
D_A = \partial_A - iA_A~, \qquad \qquad  A_A = A_A^aQ^a ,
\eeq
where $Q^a$ is an abstract operator that takes on
the values of $Q^a=\lambda^a/2$ in the adjoint representation.
The field strength then is 
\beq
G_{AB} = i[D_A,D_B]
=\partial_A A_B-\partial_B A_A - i[A_A,A_B] .
\eeq
This theory has the standard kinetic term:
\beq
\label{start}
{\cal{L}} = -\frac{1}{2\tilde{g}^2}\Tr(G_{AB}G^{AB})  ,
\eeq
where $1/\tilde{g}_2^2$ is the coupling with dimensions of mass. With this
normalization, gauge fields have the canonical dimensionality 
with respect to $D=4$, \ie, $[A_A]= M^{1}$, and $[G_{AB}]= M^2$. 

The theory possesses two identically conserved Chern-Simons 
currents of the form:
\bea
\label{cur1}
J_A & = & \epsilon_{ABCDE}\Tr(G^{BC}G^{DE}),
\eea
\bea
\label{current2}
J^a_A & = & \epsilon_{ABCDE}
\Tr \Bigl ( \frac{\lambda^a}{2}\{G^{BC},G^{DE}\}\Bigr ).
\eea
The second current requires that $SU(N)$ possess a $d$-symbol, 
hence $N\geq 3$; and it is further {\em covariantly}
conserved, 
\beq
[D^A,~J^a_A~ Q^a]=0. 
\eeq
These topological  
currents do not arise from eq. (\ref{start}) 
under local Noetherian variation of the fields. 

The adjoint currents can be derived from an
action containing the ``second Chern character.''
The second Chern character, which we'll abbreviate as CS2, 
is derived by ascending to 
$D=6$ and considering the generalization of the Pontryagin index
(a $D=6$ generalization of the $\theta$-term),
\beq
{\cal{L}}_0 =\epsilon_{ABCDEF}\Tr G^{AB}G^{CD}G^{EF} .
\eeq
This can be written as a total
divergence, 
\beq
\frac{1}{8}{\cal{L}}_0 
=
-\partial^F\epsilon_{ABCDEF}
\Tr\Bigl (A_A \partial_B A_C \partial_D A_E   
- \frac{3i}{2}A_A A_BA_C \partial_D A_E 
- \frac{3}{5}A_A A_B A_C A_D A_E\Bigr ). 
\eeq
Formally, compactifying the sixth
dimension and integrating ${\cal{L}}_0$ over 
the boundary in $x^5$ leads to ${\cal{L}}_1$,
the second Chern character as an element
of the $D=5$ Lagrangian, 
\bea
\label{CSterm0}
{\cal{L}}_1 & = & c\epsilon^{ABCDE}
\Tr \Bigl (A_A \partial_B A_C \partial_D A_E   
 - \frac{3i}{2}A_A A_BA_C \partial_D A_E 
- \frac{3}{5}A_A A_B A_C A_D A_E \Bigr ) .
\eea
This can be rewritten in a convenient
form involving gauge covariant field
strengths,
\bea
\label{CS2}
{\cal{L}}_1 
& = &
\frac{c}{4}\epsilon^{ABCDE}
\Tr \Bigl (A_A G_{BC}G_{DE}   
 + i A_A A_B A_C G_{DE}
- \frac{2}{5}A_A A_B A_C A_D A_E \Bigr ) ,
\eea
hence, for pure gauge configurations all but the last term vanish. 
The second Chern characters can be constructed in any odd
dimension from a general algorithm \cite{Wu}.

While not manifestly gauge invariant, it is
straightforward to verify that CS2 is indeed gauge
invariant for gauge transformations continuously
connected to the identity. By contrast, for topologically
nontrivial gauge transformations, the action shifts by a constant. 
Hence, the coefficient $c$ must be chosen
for effective invariance so that the action shifts by $2\pi N$: 
the path integral is then invariant. 
It can be shown that this factor is:
\beq
c = \frac{1}{48\pi^2} ~.
\eeq
The
variation of the action with respect to the gauge field $A^a$ indeed 
generates the current of eq.(\ref{current2}) as a source for the
equation of motion.

\newpage
\vskip 0.15in
\noindent{\bf (ii) Heuristic Derivation of $D=4$
Wess-Zumino-Witten Term}
\vskip 0.15in

Consider orbifold compactification of the $D=5$ Yang-Mills theory
to a $D=4$ theory.
Orbifold compactification is usually specified mathematically
following Horava and Witten \cite{horava}, such as
``compactification on $S_1/Z_2$.'' One thus
considers an interval $0\leq x^4\leq 2a$, 
classifies basis functions as even
$P=(+)$ or odd $P=(-)$ under reflection about $x^4=a$, and under
compactification, demands that the $P=(+)$ basis functions 
are assigned to the 
$D=4$ vector potentials, $A_\mu^A$, and the $P=(-)$ basis functions
to the $A_4^A$ vector potentials. 
Orbifolding is the basis of many
models of low energy extra dimensions, but we prefer a more
physical statement on orbifold compactification. 

Alternatively, we can consider two branes to be located at
$x^4=0$ and $x^4=a$. Each brane $i$ has a normal vector 
$\eta_i^A$; \eg, for brane ``L'' we have $\eta_L=(0,0,0,0,1)$
and for brane ``R'' we have $\eta_{R} = (0,0,0,0,-1)$. The orbifold
boundary conditions can be viewed as a special gauge choice
for the boundary condition applied on each brane of: 
\beq
 \left. \eta_{A} ~G^{AB} \right|_{L,R} = 0.
\eeq
This boundary condition is manifestly gauge invariant. 
For the $\eta_i$ defined above, we see that $G_{04}=0$,
hence the normal component
of the chromoelectric field strength is zero. Moreover,
the ``parallel'' magnetic field $G_{\mu 4} $ where $\mu \neq 0$
is also zero. These boundary conditions on the
branes are dual to those of an electric superconductor, and
they thus correspond to a magnetic superconductor. A magnetic
superconductor would form electric flux tubes between electric charges
(quarks) in the medium, hence a magnetic superconducting phase is
a confinement phase.

We thus consider the orbifold compactification as a kind of parallel
plate magnetic superconducting capacitor (it can likewise be 
viewed as a magnetic superconducting
Josephson junction). Spanning the gap
between the plates,  is a Wilson line:
\beq
U = P\exp\left( -i\int_0^{a} dx^A A_A \right) = \exp(i\tilde{\pi}),
\eeq
where, upon compactification, we view the Wilson line as a
chiral field of mesons, as indicated, with $\tilde{\pi} =
\pi^a\lambda^a/f_\pi $, where $f_\pi = 95$ MeV. 

In the superconducting boundary brane, or
capacitor plate regions (we'll refer to these generically
as the ``end-zones''), we 
can perform local gauge transformations. If the gauge group is $SU(N)$,
then there exist gauge transformations $V_L$ ($V_R$) that are
constant over the entire left-hand (right-hand) end-zone. These
can be identified as global $SU(N)_L$ ($SU(N)_R$) transformations. Under
these transformations we see that $U$ transforms as 
\beq
U \rightarrow V_L U V^\dagger_R ~, 
\eeq
and the theory under compactification becomes a gauged $SU(N)_L\times
SU(N)_R$ chiral lagrangian. The gauge fields should be viewed
as left-- and right-- handed combinations of
the normal vector and axial vector mesons of QCD, and
they should be supplemented with additional Higgs fields
to acquire masses. We thus do not pass to a unitary gauge
in which the $A_4$ modes are eaten by gauge fields to acquire masses.

In the end-zones, we have
the magnetic
superconducting phase. Here we hypothesize that the vector potentials
are determined by ``London currents,'' the chiral currents
built out of the Wilson line:
\beq
\label{london}
A_{A}(\makebox{L end-zone}) = iU[\partial_A, U^\dagger] \equiv i\alpha_A  ~,
\qquad
A_{A}(\makebox{R end-zone}) = iU^\dagger[\partial_A, U] \equiv i\beta_A ~.
\eeq
London currents are generated by the magnetic condensate kinetic term (\eg,
analogous to a Higgs field), that locks the vector
potential to the Nambu-Goldstone boson, \eg, in our
present case $A_A(L)= \partial_A\tilde{\pi} + ...$ 
and $A_A(R)= -\partial_A\tilde{\pi} + ...$ in the endzones.
The particular definitions given in eq.\ (\ref{london})
are pure gauges, and thus the gauge field
strength vanishes (\eg, using form notation,
$d\alpha = -\alpha^2$, and   $(1/2)G = dA-iA^2 = 0$ when $A=i\alpha$).

We now seek the low energy effective 
theory. We substitute the London current vector
potentials into the Chern-Simons term of eq.(\ref{CS2})
to obtain the $D=4$ effective topological
 lagrangian:
\beq
 \left(\frac{1}{2\times 5}\right) \frac{i}{48\pi^2}\epsilon_{ABCDE}\left(
 \Tr\alpha^A\alpha^B\alpha^C\alpha^D\alpha^E +  
 \Tr\beta^A\beta^B\beta^C\beta^D\beta^E\right).
\eeq
where the $\alpha$ ($\beta$) terms reside on the left (right)
end-zone.
To leading order in the expansion
in pions, we can write, 
\bea
\epsilon_{ABCDE}\Tr\alpha^A\alpha^B\alpha^C\alpha^D\alpha^E & = &
i\epsilon_{ABCDE}\partial_A\Tr\tilde{\pi}\alpha^B\alpha^C\alpha^D\alpha^E + ...,
\nonumber \\
\epsilon_{ABCDE}\Tr\beta^A\beta^B\beta^C\beta^D\beta^E & = &
-i\epsilon_{ABCDE}\partial_A\Tr\tilde{\pi}\beta^B\beta^C\beta^D\beta^E + ...,
\eea
Thus, when we integrate 
$x^4$ over
the gap between the end-zones, $\int_0^a dx^4 $,
we arrive at the effective lagrangian, 
\beq
\label{WZ}
 \frac{1}{240\pi^2}\epsilon_{\mu\nu\rho\sigma}\left(
 \Tr\tilde{\pi}\alpha^\mu\alpha^\nu\alpha^\rho\alpha^\sigma \right).
\eeq
Eq.(\ref{WZ}) is the precise structure of the leading
piece of the Wess-Zumino term in an expansion 
in pions with
Witten's normalization. 

A few comments are in order. Note that the expression
is hermitian---and it can be written as either
$\Tr(\pi \alpha^4)$  or $\Tr(\pi \beta^4)$. Witten's derivation 
involves compactification on a disk, and the WZW term
resides on the periodic boundary of the disk, while
the present approach has used the orbifold configuration. 
Witten writes in an expansion in pions $(2/15\pi^2F_\pi^5)\Tr(A\partial_\mu A
\partial_\nu A\partial_\rho A\partial_\sigma A) + ...$
with $A=\pi^a\lambda^a$ and $F_\pi = 2f_\pi$, which is consistent
with eq.(\ref{WZ}). We note that the $\alpha$ terms 
in the above derivation received a minus sign upon integrating
from $0$ to $a$
(which canceled against $i^2$) since the left end-zone
resides at the lower limit of the integral; the $\beta$
terms received a positive sign. In
the $D=4$  theory the currents $\alpha(x_\mu)$ and $\beta(x_\mu)$ 
are viewed as residing
at a common point in $D=4$ space-time, and we then have
identities such as: 
\beq
\epsilon_{ABCDE}\Tr\tilde{\pi}\beta^B\beta^C\beta^D\beta^E
=\epsilon_{ABCDE}\Tr U\tilde{\pi}U^\dagger\alpha^B\alpha^C\alpha^D\alpha^E,
\eeq
and $U\tilde{\pi}U^\dagger =\tilde{\pi}$, and 
we use $U\beta U^\dagger = -\alpha$
and $U\tilde{\pi} U^\dagger$ to bring the two terms into the
common form.

With covariant London currents, \eg,
$\alpha_A \rightarrow U[D_A, U^\dagger]$, the expression
 becomes fully gauge invariant. The field strength is then
 nonzero, and other
operators like $\Tr(\pi \alpha^2 G)$, $\Tr(\pi \alpha G \alpha)$, \etc,
now appear. This expression can be integrated by parts
into the full Wess-Zumino-Witten term 
which will be developed
in greater detail elsewhere \cite{hill}.

The present ``derivation'' is only meant to be heuristic, and
is not well-defined (\eg, operators like $ U[D_A, U^\dagger]$
have path dependence). Nonetheless, there
are many alternative ways to proceed to formalize the deformation
theory from $D=5$ pure Yang-Mills into $D=4$ chiral lagrangians. 
In the subsequent sections we'll
be led to a particular and well-defined deformation 
of the $D=5$ Yang-Mills theory
into a $D=4$ chiral lagrangian in which, \eg, $A_\mu(L)\rightarrow
A_\mu(L)+i\half U[D_\mu, U^\dagger]$. 

\vskip 0.15in
\noindent{\bf (iii) Singlet Auxiliary Chern-Simons Term
and a New Singlet WZW Term}
\vskip 0.15in

We presently turn to the singlet topological current,
and we'll merely anticipate some results that follow for
the compactification and deconstruction, using the
techniques of the next section. 
  
The singlet Chern-Simons current can be
generated by an additional modification 
of the Lagrangian of the form
(CS1):
\beq
\label{CS1}
{\cal{L}}_2 = {c'}\epsilon_{ABCDE}V^A\Tr(G^{BC}G^{DE}),
\eeq
where $V^A$ is a singlet auxiliary vector field. Since it is identically 
conserved, the CS singlet current couples to this vector field 
in ${\cal{L}}_2$ compatibly with a simple abelian gauge-invariance, 
$\delta V^A= \partial_A \sigma$. 
If the vector field $V$ is endowed with kinetic terms, the singlet current 
is also generated as a source in the corresponding Maxwell equations
of motion for $V$.  
Note that the singlet current cannot be derived
from CS2, as the Chern-Simons term of eq.(\ref{CSterm0}) only exists
in $SU(N)$ for $N\geq 3$ and does not occur, e.g., in $SU(2)$ Yang-Mills, while
the singlet current is always present. One might argue
that in $SU(2)$ the form of the current can be inferred, {\em a posteriori},
e.g., by considering the $\lambda^8$
component in $SU(3)$ of the adjoint current, and setting coset fields
to zero to descend to $SU(2)$. The singlet current cannot
arise from direct variation of CS2, eq.(\ref{CSterm0}), and 
eq.(\ref{CS1}) (CS1) is required to generate it {\em a priori}. 

The appearance of $V_A$ is linked to the instantonic soliton 
\cite{topo1,topo2}, the 't Hooft 
instanton lifted to a static ``monopole'' configuration in $D=5$.
This object has  a mass of $8\pi^2/\tilde{g}^2$, and it 
descends to the 
skyrmion, characterized by the Goldstone-Wilczek current \cite{goldstone},
in $D=4$
\cite{topo2}.                  
$V_A$ can be interpreted as an effective field associated with the 
instantonic soliton. 
The choice of $V_A$ is dictated by the degrees of freedom in the theory. 
We must generate a conserved current, hence the
variation $\delta V^A= \partial_A \sigma $, \ie, we have
no complex fields to draw upon. However, the instantonic
soliton must be described as a massive excitation, hence we cannot use
a Nambu-Goldstone field $\sigma$ by itself. We may thus infer that the
instantonic soliton is associated with a massive $U(1)$ gauge field.

Making use of the chiral deconstruction techniques
discussed in the present paper, we can deconstruct
eq.(\ref{CS1}) to obtain a new auxiliary WZW term that generates
the Goldstone-Wilczek current. The field $V^A$ is
decomposed into $V^A_L+V^B_R$ where $V_L$ ($V_R$) has support
in the $L$ ($R$) end-zone.
The $x^4$ integrated zero modes of $V^A$
are then defined in terms of $\sigma$ and $\eta'$ fields
of a chiral theory of mesons:
\bea
\int dx^4 V_{4R} & = & a(\sigma + \eta') ,
\quad
\int dx^4 V_{4L}  = a(\sigma - \eta') , \nonumber \\
\int dx^4 V_{\mu R} & = & af^{-1}\partial_\mu(\sigma - \eta' ),
\qquad
\int dx^4 V_{\mu L} = af^{-1}\partial_\mu(\sigma + \eta'),
\eea
where the choices are consistent with
parities, and the Noether variations
that we would make for the original $V^A$
to generate the currents (here we have set
the decay constant of $\sigma$ and $\eta'$
to unity). Note that $\sigma$ and $\eta'$
can be viewed as glueballs, physical objects
in the end-zone phases, even though the
theory is quarkless.
We find  that CS1, using methods
developed in the next section, deconstructs 
to terms containing the following form:
\bea
\label{CS2GW04}
{\cal{L}}_2 
& \rightarrow & -\frac{ac'}{2} \eta'\epsilon_{\mu\nu\rho\sigma}
\Tr\left(G_L^{\mu\nu}G_L^{\rho\sigma} 
+ G_R^{\mu\nu}G_R^{\rho\sigma}
\right) 
+i\frac{ac'}{2f} \partial^\mu (\eta') \epsilon_{\mu\nu\rho\sigma}
\Tr\left(\alpha^\nu  G_L^{\rho\sigma}
-\alpha^\nu  \overline{G}_R^{\rho\sigma}\right)  \cr
&& \cr
&&
-\frac{2ac'}{f}\partial^\mu \sigma \epsilon_{\mu\nu\rho\sigma}
\Tr\left(\frac{3 i}{2} \alpha^\nu  G_L^{\rho\sigma}
+\frac{3i}{2} \alpha^\nu  \overline{G}_R^{\rho\sigma}
+ \alpha^\nu \alpha^\rho \alpha^\nu \right)
\cr
&&
\qquad \qquad \qquad \qquad \qquad \qquad
+\;\;\frac{3a}{2}c'\sigma \epsilon_{\mu\nu\rho\sigma}\Tr\left(
G_{L\mu\nu}G_{L\rho\sigma} - G_{R\mu\nu} {G}_{R\rho\sigma} 
\right),  \cr
&&
\eea
where $\alpha = U[D, U^\dagger]$ and $\beta=U^\dagger[D, U]$,
and $\overline{G}_R^{\rho\sigma} = U{G}_R^{\rho\sigma}U^\dagger$.
This is a new WZW-like term
that correctly generates the full Goldstone-Wilczek current,
\cite{topo2,goldstone} 
with the correct normalization of a unit of baryon number for the
skyrmion, provided $c' = 1/48\pi^2$ (identifying $c=c'$
of the second Chern character), 
\be
\label{eqb}
{Q}^\mu = \frac{1}{24\pi^2}\epsilon^{\mu\nu\rho\sigma}\Tr \left(  
\alpha_\nu \alpha_\rho \alpha_\sigma  
+\frac{3i}{2}(
 G_{L\nu\rho} \alpha_\sigma + \overline{G}_{R\nu\rho} \alpha_\sigma)\right).
\ee
Thus, by constructing the Noether equation
of motion of the $\sigma$ meson,
we generate
the full conservation equation of the GW current, including its anomaly,
\beq
\partial_\mu {Q}^\mu = 
-\frac{1}{32\pi^2}\epsilon^{\mu\nu\rho\sigma}\Tr \left(  
G_{L\mu\nu}G_{L\rho\sigma}  - {G}_{R\mu\nu}{G}_{R\rho\sigma} \right).
\eeq
This shows that the singlet topological Chern-Simons current matches
the full GW current under compactification. 

Moreover, by Noether variation of
the $\eta'$, we obtain a ``$U(1)$ axial current,'' 
\be
\label{cur3}
Q_5^\mu  = \frac{Z}{32\pi^2}\epsilon^{\mu\nu\rho\sigma}\Tr \left(  
G_{L\nu\rho} \alpha_\sigma - \overline{G}_{R\nu\rho} \alpha_\sigma  
 )\right),
\ee
This actually has an indeterminate normalization
$Z$. Its divergence equation likewise follows from
the $\eta'$ equation of motion:
\bea
\label{eqb}
\partial_\mu {Q}_5^\mu  & = & \frac{Z}{32\pi^2}
\epsilon^{\mu\nu\rho\sigma}
\Tr\left[
iG_{L\mu\nu }\alpha_\rho\alpha_\sigma
+i\overline{G}_{R\mu\nu }\alpha_\rho\alpha_\sigma
+G_{L\mu\nu }G_{L\rho\sigma} + {G}_{R\mu\nu }{G}_{R\rho\sigma}
\right],
\eea
and $Z=1$ is thus favored by matching to the $U(1)$ axial anomaly.
The last two terms are the correct form of an
axial current anomaly, while the first terms on
the {\em rhs} are 
analogous to the Skyrme terms. The first two terms
form a pseudoscalar and can be
interpreted as $2im\bar{\psi}\gamma^5\psi$
in the axial current divergence of a massive nucleon.
This current, to our knowledge, has not been
previously discussed in the literature.
The details of this derivation will be 
presented elsewhere, \cite{hill}.

\vskip 0.15in
\noindent
\section{Deconstruction and Bianchi Identities}
\vskip 0.15in

The heuristic argument presented in section II
suggests that a direct morphing of 
the Chern-Simons terms  of $D=5$ Yang-Mills into $D=4$ chiral
lagrangians is possible and meaningful. We expect that
there are many possible deformations of the parent theory
in $D=5$, through deconstruction, that can yield various
chiral theories in $D=4$. These deformations may or may not
exploit the full geometrical and topological matching. We 
would expect that
an integral multiple of
the minimal coefficient of Witten, $1/240\pi^2$, will always obtain in
a consistent theory. 

The heuristic argument indeed gave the ``minimal
coefficient'' of the WZ term of Witten.  We will now turn to a more
literal interpretation of dimensional deconstruction of
pure $D=5$ Yang-Mills which pays
closer attention to the details of topological mapping---in particular,
to the definition of motion (``hopping'') in the fifth dimension
and to the Bianchi identities.  Remarkably, the present construction
yields the WZW term with a coefficient that is of the form $N/240\pi^2$,
where the index $N=D=5$, is the dimensionality of the parent space-time.

\vskip 0.15in
\noindent{\bf (i) Preliminaries}
\vskip 0.15in

We presently consider the 
compactification of the $SU(N)$ Yang-Mills theory in $D=5$
on the interval $0\leq x^4 \leq a$. First, construct
a coarse-grained lattice of the $x^4$ dimension
with $2$ slices.  
On each slice lies a copy of the gauge group with hermitian generators 
$Q_i^a$. The covariant derivative is a sum over all slices with
the appropriate abstract charge assigned to each gauge field,
\beq
D_\mu = \partial_\mu -i A_{L\mu}^a Q_L^a - i A_{R\mu}^a Q_R^a,
\eeq
where we use the notation ``left,'' $L$ (``right,'' $R$) 
for brane 1 (2).
The generators $Q_L$ and $Q_R$ act on
the given slice, and the 
slices are connected from $L$ to $R$ 
by a unitary Wilson link $U$,
connecting the 1st to the 2nd slice (while the link $U^\dagger$ 
connects slice 2 to slice 1). Thus,
\beq
[Q_L, Q_R ] = 0 .
\eeq
The hermitian field
strength tensor is,
\beq
G_{\mu\nu} = i[D_\mu, D_\nu] = G_{L\mu\nu}^a Q^a_L + G_{R\mu\nu}^a Q^a_R ,
\eeq
and also resolves into $L$ and $R$ operator components.

\vskip 0.15in
\noindent{\bf (ii) Matrix Formalism}
\vskip 0.15in

We choose to define a ``left-handed derivative,'' 
$D_{L\mu} = \partial_\mu -i A_{L\mu}^a Q_L^a$ , 
so that $G_{L\mu\nu}^a Q_L^a = i[D_{L\mu}, D_{L,\nu}]$;  
and, respectively, a ``right-handed derivative,'' 
$D_{R\mu} = \partial_\mu -i A_{R\mu}^a Q_R^a$,
so that $G_{R\mu\nu}^a Q_R^a = i[D_{R\mu}, D_{R,\nu}]$
for the right-handed fields. $D_L$ applies to fields
on the left-hand lattice slice, while $D_R$
applies on the right-hand slice. We further 
require $[D_{L\mu}, D_{R\nu }] =0$, which
does not hold, naively;  however, we can still implement
this construction as a $2\times 2$ matrix representation.

Operators are defined as left-handed and right-handed 
in the chirality matrix format,
\beq
{\cal{O}} 
= \left( \begin{array}{cc} 
O^L & 0 \\
0 & O^R \\
\end{array} \right).
\eeq
Hence, the matrix covariant derivative can be defined
as
\beq
{\cal{D}}_\mu 
= \left( \begin{array}{cc} 
D_\mu^L & 0 \\
0 & D_\mu^R \\
\end{array} \right).
\eeq
The commutator, then, yields the field strengths residing
on their respective lattice slices:
\beq
{\cal{G}}_{\mu\nu} = i[{\cal{D}}_\mu, {\cal{D}}_\nu]  
= \left( \begin{array}{cc} 
G^L_{\mu\nu} & 0 \\
0 & G^R_{\mu\nu} \\
\end{array} \right).
\eeq
The gauge transformations in this space are thus,
\beq
{\cal O} \rightarrow {\cal V O V}^\dagger ,
\qquad
{\cal V} = \left( \begin{array}{cc} 
V_L & 0 \\
0 & V_R \\
\end{array} \right).
\eeq
Lattice link fields are off-diagonal matrices,
\beq
{\cal{U}} = \left( \begin{array}{cc} 
0 & U \\
0 & 0 \\
\end{array} \right), 
\qquad
{\cal{U}}^\dagger = \left( \begin{array}{cc} 
0 & 0 \\
U^\dagger & 0 \\
\end{array} \right).
\eeq
Note, then, 
\beq
{\cal{U}}^\dagger {\cal{U}} = \left( \begin{array}{cc} 
0 & 0 \\
0 & 1 \\
\end{array} \right),
\qquad
{\cal{U}} {\cal{U}}^\dagger = \left( \begin{array}{cc} 
1 & 0 \\
0 & 0 \\
\end{array} \right),
\eeq
so that 
\beq
{\cal{U}} {\cal{U}}^\dagger + {\cal{U}}^\dagger {\cal{U}} = \openone ~,
\qquad
{\cal{U}} {\cal{U}}^\dagger - {\cal{U}}^\dagger {\cal{U}} = \sigma_z .
\eeq

The lattice Wilson links transform as bifundamentals,
\beq
{\cal{U}} \rightarrow {\cal  V U V}^\dagger = \left( \begin{array}{cc} 
0 & V_L U V_R^\dagger \\
0 & 0 \\
\end{array} \right),
\qquad
{\cal{U}}^\dagger \rightarrow {\cal V {U}}^\dagger {\cal V}^\dagger 
= \left( \begin{array}{cc} 
0 & 0 \\
V_R U^\dagger V_L^\dagger & 0 \\
\end{array} \right).
\eeq
The commutators of operators with link fields are:
\bea
\label{comm}
[{\cal{O}}, {\cal{U}}] & = & \left( \begin{array}{cc} 
0 & O_LU-UO_R \\
0 & 0 \\
\end{array} \right),
\qquad
[{\cal{O}}, {\cal{U}}^\dagger]  =  \left( \begin{array}{cc} 
0 & 0 \\
U^\dagger O_L - O_R U^\dagger & 0 \\
\end{array} \right).
\eea
The abstract charge is defined as
\beq
{\cal Q}^a = \left( \begin{array}{cc} 
Q^a_L & 0 \\
0 & Q^a_R \\
\end{array} \right).
\eeq

Thus, define the $Q^a$ as having commutators on the $U$'s:
\beq
T^a \equiv  \frac{\lambda^a}{2} ,
\qquad
[Q_L, U] = T^aU ,\qquad 
[Q_R, U] =  -UT^a .
\eeq
We often enconter these charges sandwiched between $U$ and
$U^\dagger$ matrices. We thus see that, e.g.,
\bea
U^\dagger Q_L^a U & = & U^\dagger T^a U  + Q_L^a .
\eea

The structure of eq. (\ref{comm}) allows 
covariant differentiation to be written as a commutation
relation, and takes the following form on $U$, 
\beq
[D_\mu , U] = \partial_\mu U -iA_{L\mu}^a\frac{\lambda^a}{2}U
+ iA_{R\mu}^aU\frac{\lambda^a}{2} .
\eeq
This corresponds to the chirality matrix commutator, 
\beq
[{\cal{D}}_\mu, {\cal {U}}]
= \left( \begin{array}{cc} 
0 & D_{L\mu}U-UD_{R\mu} \\
0 & 0 \\
\end{array} \right).
\eeq

From the link field $U$, we may thus form left-handed (right-invariant), 
and right-handed 
(left-invariant) chiral currents, respectively (non-matrix),
\beq
\label{alpha}
\alpha_{\mu} \equiv U [D_\mu , U^\dagger],  
\qquad \qquad 
\beta_{\mu} \equiv  U^\dagger [D_\mu , U] .
\eeq
More explicitly, 
\bea
\alpha_\mu & = & 
 U(\partial_\mu - iA^a_{R\mu}\frac{\lambda^a}{2} )U^\dagger 
 + iA^a_{L\mu}\frac{\lambda^a}{2} 
 = U(D_{R\mu}U^\dagger -U^\dagger D_{L\mu}),
\cr
\beta_\mu & = &  U^\dagger(\partial_\mu - iA^a_{L\mu}\frac{\lambda^a}{2} )U 
 + iA^a_{R\mu}\frac{\lambda^a}{2} 
 = U^\dagger(D_{L\mu}U -U D_{R\mu}),
\eea
where the action of the derivatives follows Leibniz's rule, 
$~D_L U = [D_L,U] + UD_L$; likewise, 
$D_R U^\dagger = [D_R,U^\dagger] + U^\dagger D_R$.

In the chiral matrix representation, these amount to
\beq
{\hat{\alpha}}_{\mu} = {\cal{U}}[{\cal{D}}_\mu, {\cal{U}}^\dagger]  
= \left( \begin{array}{cc} 
\alpha_{\mu} & 0 \\
0 & 0 \\
\end{array} \right),
\qquad
{\hat{\beta}}_{\mu} = {\cal{U}}^\dagger[{\cal{D}}_\mu, {\cal{U}}]  
= \left( \begin{array}{cc} 
0 & 0 \\
0 & \beta_{\mu} \\
\end{array} \right).
\eeq

Finally, it is useful to define the hermitian link chiral matrix:
\beq
{\cal{U}}_+ \equiv  {\cal{U}} + {\cal{U}}^\dagger 
= \left( \begin{array}{cc} 
0 & U \\
U^\dagger & 0 \\
\end{array} \right).
\eeq
Thus,
\beq
{\cal{U}}_+ {\cal{U}}_+ = \openone ,
 \eeq
and one sees that 
\beq
\label{central}
{\cal{A}}_\mu \equiv {\hat{\alpha}}_{\mu}+ {\hat{\beta}}_{\mu}      = 
{\cal {U}}[{\cal {D}}_\mu, {\cal {U}}^\dagger]  + {\cal {U}}^\dagger
[{\cal{D}}_\mu, {\cal {U}}]
=  {\cal{U}}_+[{\cal D}_\mu, {\cal{U}}_+ ]
=
\left( \begin{array}{cc} 
\alpha_{\mu} & 0 \\
0 & \beta_\mu \\
\end{array} \right).
\qquad
\eeq
A useful set of relationships that recur throughout,
especially in computing current divergences,
is 
\bea
\label{alpha1}
[ D_\mu, \alpha_\nu ] - [D_\nu, \alpha_\mu]
& = &
-[\alpha_\mu, \alpha_\nu] -iU[G_{\mu\nu}, U^\dagger], \cr
[D_\mu, \beta_\nu] - [D_\nu, \beta_\mu]
& = &
-[\beta_\mu, \beta_\nu] -iU^\dagger[G_{\mu\nu}, U] ,
\eea
with the correspondence in the chirality matrix representation:
\bea
\label{curvature}
[{\cal{D}}_\mu, {\cal{A}}_\nu]
-[{\cal{D}}_\nu, {\cal{A}}_\mu]
& = & - [{\cal{A}}_\mu, {\cal{A}}_\mu] -
i {\cal{U}}_+ [{\cal{G}}_{\mu\nu}, {\cal{U}}_+ ]
\cr &&\cr
& = & \left( \begin{array}{cc} 
iG^L_{\mu\nu} -iU G^R_{\mu\nu}U^\dagger -[\alpha_{\mu},\alpha_\nu] & 0 \\
0 & 
iG^R_{\mu\nu} -iU^\dagger  G^L_{\mu\nu}U-[\beta_{\mu},\beta_\nu] \\
\end{array} \right).
\eea

How should ${\cal D}_4$ be defined in D=4? 
${\cal D}_4$ is some kind of a lattice 
derivative, or ``brane-hop'', in the $x^4$ direction. In general, hopping 
on an $N$-slice lattice works through the Wilson link, $U_i$, fields,
which are identified with the continuum $A^4$ through:
\beq
U_i = P \exp \left( -i\int_{x^4_i}^{x^4_{i+1}} A_4 dx^4 \right ) .
\eeq 
Consider a field $\psi_i(x)$ on the $i$th slice, where $\psi_i(x)
\rightarrow V_i(x) \psi_i (x)$ under local gauge transformations
$V_i(x)$ of the local gauge symmetry group on the $i$th slice.
To define a covariant derivative in the $x^4$ direction, 
one seeks a difference like $(\psi_{i+1}(x) - \psi_i(x))/a$, for lattice 
spacing $a$. But since this has a mixed gauge symmetry, 
one is led to define the covariant difference 
$(U_{i}\psi_{i+1}(x) - \psi_i(x))/a$.
The link now pulls the first term back from the slice
$i+1$ to $i$, where the $i$-covariant difference can be computed,
invariant under $i+1$ transformations. A vanishing covariant difference 
thus amounts to link-gauge transformation. For adjoint quantities, both sides 
of the corresponding operator need such adjustment.

The hopping derivative in deconstruction must handle left and right in 
a manner consistent with parity. One possibility would be to define an 
{\em off diagonal} (antihermitian) hopping derivative as a commutator,
and thus traceless,
\beq
[{\cal  D}^4, {\cal  O}] \equiv -\frac{1}{a} [{\cal{U}}_+,{\cal{O}}]=
-\frac{1}{a} \left( \begin{array}{cc} 
0 & -UO_R + O_L U \\
O_RU^\dagger - U^\dagger O_L & 
0
\end{array} \right).
\eeq
With this definition, the hopping derivative obeys Leibniz's chain rule 
of differentiation, as a commutator, and so the Bianchi identities in $D=4$ 
are automatically satisfied, so there is no need for modification of the 
theory. Thus, an orbifold
compactification solves the Bianchi identity with the usual spectrum.
In addition, by Leibniz's rule, ${\cal O \psi}$ hop-transforms 
exactly like ${\cal \psi}$.  This may be at the root of a deficiency,
however. Being off diagonal, this ${\cal D}_4$ maps operators from one
representation into another. For example,
it maps an adjoint representation under $SU(N)_L$, \ie, $(N^2-1, 0)$,
into  a bifundamental under $SU(N)_L\times SU(N)_R$, \ie $(N,\bar{N})$.
A covariant derivative then which does not faithfully map a given
representation into itself is unsatisfactory. 

Moreover, if it 
is applied to fermions, one immediately encounters the fermion doubling 
problem. The remedy
to this is the addition of a Wilson term, which is a 
continuum second derivative.
If we generalize the Wilson term to the case of
higher representations, such as adjoint
operators, we are led to the diagonal definition, below \cite{leib}.
The Wilson term projects out the unwanted fermionic doublers, and 
permits the appearance of anomalies consistently with topology. 
Multiplication of the above ${\cal D}_4$ by $-\sigma_z {\cal{U}}_+ $  
on the left, however, leads to a different, {\em diagonal hopping derivative} 
defined below.

A preferred definition, and the one we will be using presently is a 
{\em diagonal hopping derivative},
\beq
{\cal D}^4 ({\cal O}) \equiv 
\frac{1}{a}\left({\cal{U}}[{\cal{O}},{\cal{U}}^\dagger]
-{\cal{U}}^\dagger[{\cal{O}},{\cal{U}}]\right)
=
\left( \begin{array}{cc} 
UO_RU^\dagger - O_L  & 0 \\
0 & O_R - U^\dagger O_L U
\end{array} \right),
\eeq
where $ a$ is the spacing between neighboring slices.

Note the second term pushes the previous slice fields forward,
as the first term pulls the subsequent slice fields back, hence
a relative sign difference, which is commensurate 
with parity:
 Under parity, $L\leftrightarrow R$,
$U\leftrightarrow U^\dagger$,
and ${\cal D}^4 \rightarrow -{\cal D}^4$, and hence the definitions are
parity invariant.  It is important to note, however, that, like lattice 
derivatives, this derivative does not obey the Leibniz rule of differentiation,
and so {\em cannot} be written as a commutator 
$[{\cal D}^4 , {\cal A} {\cal B}]= 
[{\cal D}^4 , {\cal A}] {\cal B}+ {\cal A} [{\cal D}^4 ,  {\cal B}]$.

We thus define the coset field strength as a transform, not a commutator,
through the diagonal hopping derivative:
\bea
\label{trivium}
{\cal G}_{4\mu} & =& -{\cal G}_{\mu 4} \equiv i{\cal D}_4 ( {\cal D}_\mu)  
\nonumber \\               & = & 
 \frac{i}{a}\left({\cal{U}} [{\cal D}_\mu, {\cal {U}}^\dagger]
- {\cal{U}}^\dagger[{\cal D}_\mu, {\cal{U}}] \right)
=\frac{i}{a}(\hat\alpha_\mu - \hat\beta_\mu) 
= \frac{i}{a}\left( \begin{array}{cc} 
\alpha_\mu & 0 \\
0 & -\beta_\mu \\
\end{array} \right).
\eea

The conventional deconstructed
lagrangian in the chirality matrix formalism can then be written,
\bea
\label{lag2}
{\cal{L}} & = & -\frac{1}{2{g}^2}\left(  
\Tr {\cal{G}}_{\mu\nu}{\cal{G}}^{\mu\nu}
-\Tr {\cal{G}}_{4\nu}{\cal{G}}^{4\nu}
\right)
\cr
& = & -\frac{1}{2{g}^2}  \Tr G_{L\mu\nu}G^{L\mu\nu}
-\frac{1}{2{g}^2}  \Tr G_{R\mu\nu}G^{R\mu\nu}
- \frac{1}{8} f_\pi^2~ \Bigl (\Tr (\alpha_\mu)^2  + \Tr (\beta_\mu)^2 \Bigr ), 
\eea
where we identify $1/g^2 = a/\tilde{g}^2$, and 
$f^2_\pi = 4/ a\tilde{g}^2 = 1/a^2{g}^2$.

It could be interpreted as a gauged chiral lagrangian
with external vector fields, $A^\mu _L$ and $A^\mu _R$. We may wish to
assign the octet of vector mesons, including the
$\rho$ to a vector combination of the fields, and the axial
vector mesons to the axial vector combination. To do
this in detail would require additional Higgs fields to give
masses to the vector ($\rho$ ) and axial vector ($A_1$)
combinations. Once these combinations have acquired with longitudinal degrees
of freedom, then one cannot eliminate the mesons by gauge transformations.

As an effective fundamental theory, this represents a massless zero mode 
together with a massive KK mode. To see this, pass to unitary gauge to remove 
the spinless mesons altogether, i.e., note that 
$\Tr (\alpha_\mu)^2 = \Tr (\beta_\mu)^2$, and introduce a ``St\"{u}ckelberg'' 
field, 
\beq
V_\mu \equiv -i\alpha_\mu/g .
\eeq
The corresponding field strength, by eq. (\ref{alpha1}) is:
\bea
F^V_{\mu\nu} & = & [D_\mu, V_\nu] - [D_\nu, V_\mu] - i[V_\mu, V_\nu]
\cr && \cr
& = & -\frac{1}{g}U[G_{\mu\nu}, U^\dagger] = 
\frac{1}{g}G^a_{L\mu\nu}\frac{\lambda^a}{2}
-\frac{1}{g}UG^a_{R\mu\nu}\frac{\lambda^a}{2}U^\dagger   .  
\eea
Further, the orthogonal zero-mode field strength is likewise right-invariant, 
\bea
F^0_{\mu\nu} & = &  \frac{1}{g}
\left(UG^a_{R\mu\nu}\frac{\lambda^a}{2}U^\dagger +
G^a_{L\mu\nu}\frac{\lambda^a}{2}\right). 
\eea  
Thus, the effective lagrangian takes the form,
\beq
\label{lag3}
{\cal{L}} = -\frac{1}{2}  \Tr F^0_{\mu\nu}F^{0\mu\nu}
-\frac{1}{2}  \Tr F^V_{\mu\nu}F^{V\mu\nu}
- \frac{1}{4}g^2f_\pi^2~ \Tr V_{\mu}V_{\mu}, 
\eeq
describing a massless zero mode and massive KK mode
of mass $gf_\pi/\sqrt{2}$.  (The spinless mesons have been
absorbed into the longitudinal components of $V_\mu$.) 

Note that one can always perform
a left gauge transformation on these fields, 
$D_R \rightarrow U D_R U^\dagger = D_R^\prime$
leading to $G^a_{R\mu\nu}{}^\prime =
UG^a_{R\mu\nu}\frac{\lambda^a}{2}U^\dagger$,
hence  $gF^V_{\mu\nu}  = G^a_{L\mu\nu}\frac{\lambda^a}{2}
-G^a_{R\mu\nu}{}^\prime\frac{\lambda^a}{2} $; thus 
$gF^0_{\mu\nu}  =  G^a_{R\mu\nu}{}^\prime\frac{\lambda^a}{2} +
G^a_{L\mu\nu}\frac{\lambda^a}{2}$. With these field
redefinitions, evidently only one linearly
realized symmetry transforms
all fields, the vectorial symmetry,
$O \rightarrow VOV^\dagger $, where $V= V_L$.

\vskip 0.15in
\noindent{\bf (iii) Bianchi Identities}
\vskip 0.15in

The {\em Bianchi identities} in $D=5$ are just the Jacobi identities for 
covariant derivatives,  
\beq
\epsilon_{ABCDE}[D^C, G^{DE}] = i \epsilon_{ABCDE}[D^C, [D^D , D^E]] =0.
\eeq
Consistency in $D=4$ requires:
\beq
\label{one}
\epsilon_{\mu\nu\rho\sigma} [{\cal D}^\nu, {\cal G}^{\rho\sigma}]=  
i\epsilon_{\mu\nu\rho\sigma} [{\cal D}^\nu, [{\cal D}^{\rho}, 
{\cal D}^{\sigma}]] 
= 0, 
\eeq
as well as,
\beq 
\label{two}
\epsilon_{\mu\nu\rho\sigma}{\cal D}^4 ( {\cal G}^{\mu\nu} )  =  
\epsilon_{\mu\nu\rho\sigma}\left([{\cal D}^{\mu}, 
{\cal G}^{4\nu}]-[{\cal D}^{\nu}, {\cal G}^{4\mu}]\right). 
\eeq
Eq. (\ref{one}) holds automatically in the $D=4$ theory,
as ${\cal G}_{\mu\nu}$ is defined as a commutator of 
covariant derivatives, for {\em any choice of ${\cal D}_\mu$}. 

The off-diagonal hopping derivative
satisfies the coset identity eq. (\ref{two}), 
while the diagonal hopping derivative, not being a commutator,  does not,
in general: the Bianchi relation implies
a nontrivial constraint. Consider the 
diagonal hopping on the {\em lhs} of eq.(\ref{two}),
\beq
\label{bianchilhs}
{\cal{D}}^4 ({\cal {G}}^{\mu\nu} )=  \frac{1}{a}\left(
{\cal{U}}[ {\cal{G}}_{\mu\nu}, {\cal{U}}^\dagger ]- 
{\cal{U}}^\dagger [{\cal{G}}_{\mu\nu}, {\cal{U}}]\right),
\eeq
and compare to the {\em rhs} of eq.(\ref{two}),
\bea
\label{bianchirhs10}
i[{\cal D}_\mu , {\cal D}_4 ( {\cal{D}}_\nu) ] 
 - i[{\cal{D}}_\nu ,  {\cal{D}}_4 ( {\cal{D}}_\mu) ] & = & 
\frac{i}{a}\left( [{\cal{D}}_\mu, {\cal{U}} 
[{\cal{D}}_\nu, {\cal{U}}^\dagger]]
-[{\cal{D}}_\mu, {\cal{U}}^\dagger [{\cal{D}}_\nu, {\cal{U}}]] - 
(\mu\leftrightarrow \nu) \right)
\nonumber \\
& = & \; \frac{i}{a}\left( {\cal{U}}[[{\cal{D}}_{\mu}, {\cal{D}}_{\nu}], 
{\cal{U}}^\dagger ]- {\cal{U}}^\dagger[[{\cal{D}}_{\mu}, 
{\cal{D}}_{\nu}], {\cal{U}} ]\right.   \nonumber \\
& & \left.
+[{\cal{D}}_{[\mu}, {\cal{U}}][{\cal{D}}_{\nu]}, {\cal{U}}^\dagger] 
- [{\cal{D}}_{[\mu}, {\cal{U}}^\dagger][{\cal{D}}_{\nu]}, {\cal{U}}] \right)
\nonumber \\
& = &
{\cal{D}}_4 ({\cal{G}}^{\mu\nu}) 
+\frac{i}{a} \left(-  [\hat{\alpha}_\mu, \hat{\alpha}_\nu] 
+ [\hat{\beta}_\mu, \hat {\beta}_\nu]\right).
\eea

The first term of the {\em rhs} is consistent,
but the last term is an unwanted nonvanishing current commutator.
This term is nonzero, and is the current algebra 
of the chiral theory. Thus, the Bianchi identity fails given
the presence of this term.

Nonetheless, the constraint, eq.(\ref{two}), can be satisfied if we
consider a {\em modified covariant derivative}.
We find that the desired modification takes the form,
\beq
\label{newderiv}
\fbox{ ${\cal{D}}'_\mu \equiv {\cal{D}}_\mu + \frac{1}{2} {\cal{A}}_\mu$} ~.
\eeq		
The Bianchi identities of eq.(\ref{one}) thus remain automatic in the 
$D=4$ subspace, since the gauge field strengths 
are defined, as usual, by commutators of ${\cal D}'_\mu$.
The Bianchi constraint, eq. (\ref{two}), 
now
requires the vanishing of the following expression,
with the modified derivative: 
\beq
\epsilon^{\mu\nu\rho\sigma} 
\left(  [{\cal{D}}'_\mu, U][{\cal{D}}'_\nu, U^\dagger  ]  
- [{\cal{D}}'_\mu, U^\dagger] [{\cal{D}}'_\nu , U] \right)
  =   0.
\eeq
To see the vanishing of this constraint, we first note:
\beq
{\cal{U}}_+ [{\cal{A}}_{\mu},{\cal{U}}_+] = -2{\cal{A}}_{\mu},
\eeq
hence, 
\beq
\{ {\cal{U}}_+ ,  {\cal{A}}_{\mu} \}=0 ,
\eeq
and so, by eqn. (\ref{central}),
\beq
[ {\cal{D}}'_{\mu}, {\cal{U}}_+] =  0.
\eeq
It is evident that this, in fact, resolves into the two components,
\beq
[ {\cal{D}}'_{\mu}, {\cal{U}}] =  0,  \qquad \qquad 
[ {\cal{D}}'_{\mu}, {\cal{U}}^\dagger ] =  0,
\eeq
and the Bianchi constraint is therefore satisfied.
We remark that one can derive the same result without recourse to the
matrix representation by careful analysis, where, allowing
an arbitrary factor $w$ in the current part
of  eq.(\ref{newderiv}), one obtains the unwanted current commutators
of eq.(\ref{bianchirhs10}), multiplied by a factor of $(1-4w+4w^2)$.
The Bianchi constraint is thus satisfied with the new 
covariant derivative
of eq.(\ref{newderiv}) with the special coefficient of $w=1/2$.
The matrix formulation both streamlines and automates this derivation.

Observe that the field strength ${\cal G}'_{4\mu}$ of (\ref{trivium}) 
manifestly vanishes for this hopping-flat modified derivative,
\beq
{\cal G}'_{4\mu} = i{\cal D}_4 ( {\cal D}'_\mu)=0.  
\eeq
(Actually, by ${\cal{D}}_4 ({\cal{G}}'_{\mu\nu})=0$, each of the three terms
in the respective coset identity eq. (\ref{two}) vanishes separately 
for modified covariant derivatives.) 

The rest of the field strength tensor, by (\ref{curvature}),
reduces to 
\beq
{\cal{G}}'_{\mu\nu} 
= i[{\cal{D}}'_\mu, {\cal{D}}'_\nu]  
=\half ({\cal{G}}_{\mu\nu}+  {\cal{U}}_+ {\cal{G}}_{\mu\nu}{\cal{U}}_+ 
-\frac{i}{2}[{\cal{A}}_\mu, {\cal{A}}_\mu] ),
\eeq
so that 
\beq
{\cal{G}}'_{\mu\nu} =
\half \left( \begin{array}{cc} 
G^L_{\mu\nu}+ U G^R_{\mu\nu} U^\dagger 
- \frac{i}{2}[\alpha_\mu,\alpha_\nu] & 0 \\
0 & G^R_{\mu\nu} + U^\dagger G^L_{\mu\nu} U
-\frac{i}{2}[\beta_\mu,\beta_\nu]\\
\end{array} \right).
\eeq
Evidently, the right-slice amounts to the gauge-transformed image 
theory of the left-slice,  
\beq
{\cal{G}}'_{\mu\nu} 
= \half \left( \begin{array}{cc} 
G^L_{\mu\nu}+ U G^R_{\mu\nu} U^\dagger 
- \frac{i}{2}[\alpha_\mu,\alpha_\nu] & 0 \\
0 & U^\dagger (U G^R_{\mu\nu} U^\dagger 
 + G^L_{\mu\nu} 
-\frac{i}{2}[\alpha_\mu,\alpha_\nu])U\\
\end{array} \right),
\eeq
For gauge-invariant combinations, this ``hop-invariant'' setup effectively 
doubles up the theory. The effective field strength appearing here 
is simply the (hop-symmetric) 
zero-mode combination encountered previously,
\beq
F_{\mu\nu}^0 = G^L_{\mu\nu}+  U G^R_{\mu\nu} U^\dagger ,
\eeq
whereas the orthogonal KK-mode combination is absent. 

In effect, the diagonal hopping derivative
Bianchi-compatible theory of the two-slice orbifold
contains only one propagating gauge field, together
with the spinless mesons. This does not mean that there is no
KK mode, but that the simplest hop-symmetric deconstruction truncates
the spectrum on the propagating zero mode. To obtain the second KK mode 
would require that we start with $N=3$ branes, and
we would expect that, for any $N$,  the Bianchi-improved theory would 
describe the zero mode and $N-2$ KK modes. 

Note that the chiral feature of an orbifold is still present, \ie, we may treat
$F_{\mu\nu}^0$ as any combination of left-hand or right-hand
gauging. For example, we may gauge only the left-hand side
of the meson fields, whence setting $ G^{R}_{\mu\nu} =0$, so that 
$F^0_{\mu\nu}= G^L_{\mu\nu}$; or, else, we may choose to gauge
isospin, $G^L_{\mu\nu} = G^{R}_{\mu\nu} $, so
$F^0_{\mu\nu}= 2G^L_{\mu\nu}$ (which rescales the coupling constant). 

In the simplifying case that we set the right-hand 
Yang-Mills fields to zero (\ie, we retain only a single
$SU(N)_L$ gauge group), we end up with a pure left-handed chiral theory:
\beq
{\cal{G}}'_{\mu\nu} = \half \left( \begin{array}{cc} 
G^L_{\mu\nu} 
- \frac{i}{2}[\alpha_\mu,\alpha_\nu] & 0 \\
0 &  U^\dagger \left( G^L_{\mu\nu} 
-\frac{i}{2}[\alpha_\mu,\alpha_\nu]\right)U\\
\end{array} \right).
\eeq
The resulting gauge action is then, 
\bea
\label{skyrmeterm}
-\frac{1}{2\tilde{g}^2}\Tr {\cal{G}}'_{\mu\nu}{\cal{G}}'^{\mu\nu}
& = & -\frac{1}{4\tilde{g}^2} 
\left(
\Tr G^L_{\mu\nu} G^{L\mu\nu} 
-i\Tr (G^{L\mu\nu} 
[\alpha_\mu, \alpha_\nu])  
- \frac{1}{4}\Tr[\alpha_\mu, \alpha_\nu][\alpha^{\mu}, \alpha^{\nu}]
 \right) . 
\eea

The resulting theory has several interesting properties 
evident at this point. The last term, $ \Tr([\alpha,\alpha]^2)$, 
is the Skyrme term required for
the stability of the core of the Skyrmion solution. 
It is normally a puzzle to understand how these terms are
generated in a deconstructed theory, since they are needed classically,
because the skyrmion core is not an entirely short-distance structure.
The Bianchi identities have fixed the coefficient of the 
Skyrme terms to definite values.
While one  could always add other contributions to the
Skyrme terms by hand, their appearance here reflects self-consistency with
the parent $D=5$ theory, which admits stable large instantonic solitons,
which, in turn, carry the current that matches to the Skyrmionic current. 

We note that the new cross-term of the form $G^L [\alpha,\alpha] $, 
which is  allowed by the presence of the
gauge field. This term has significant effects
upon the mass of the skyrmion, and bounds related to those
of magnetic monopoles arise \cite{briyahe}.

We are thus led to speculate that this Bianchi-consistent theory, 
with these fixed Skyrme terms, points to a more intricate relationship
between the instantonic soliton and the skyrmion. Perhaps we could 
now find a skyrmion solution that is ``self-dual,''  matching
the self-duality of the instantonic soliton in $D=5$, which, in turn, is
a consequence of the self-duality of the instanton. 

In non-matrix notation, the modified derivative reads,
\beq
D'_\mu = \partial_\mu -i (A_{L\mu}+ \frac{i}{2} \alpha_\mu )\cdot 
Q_L - i (A_{R\mu}+ \frac{i}{2} \beta_\mu )\cdot Q_R ,
\eeq
and hence,
\beq
\label{nonmatrixprimeder} 
D'_{L\mu}  = \half ( D_{L\mu}+ U   D_{R\mu} U^\dagger)
=  \partial_\mu - i A_{\mu L} +\half \alpha_\mu,\quad 
D'_{R\mu}  = \half ( D_{R\mu}+ U^\dagger D_{L\mu} U) 
=  \partial_\mu - i A_{\mu R} +\half \beta_\mu .
\eeq
Effectively, the gauge fields are augmented by the meson currents 
$\alpha_\mu$ and $\beta_\mu$. In the limit of vanishing gauge fields, 
the effective primed gauge fields are still non-trivial, 
\beq
\fbox{ $A'_{\mu L} \rightarrow \frac{i}{2}  U \partial_\mu U^\dagger, \qquad   
\qquad  
A'_{\mu R} \rightarrow \frac{i}{2}   U^\dagger  \partial_\mu U  $} ~,
\label{trans}
\eeq	
reminiscent of the London equation inside a superconducting medium.
Since they are not pure gauges, because of the coefficient of $1/2$, they 
yield nonvanishing primed field strengths, and hence the Skyrme term 
exhibited above. 

%

To summarize, the deconstruction prescription we have been led to 
is based on the diagonal hopping derivative 
${\cal D}_4$; the Bianchi-consistent hopping-flat 
modified covariant derivatives, ${\cal D}'_\mu$;
and the corresponding field strengths, ${\cal{G}}'_{\mu \nu}$.
Having rejected the nonvanishing ${\cal{G}}_{\mu 4}$, in favor 
of its vanishing primed counterpart, we have forfeited the meson currents' 
kinetic term, in the naive chiral lagrangian above. To recover them, we might, 
for instance, supplement the lagrangian with a term of the form:
\beq
\sim \frac{f_\pi^2 }{8}\Tr{\cal{A_\mu}}{\cal{A_\mu}},
\eeq
or somehow match ${\cal {A_\mu}}\mapsto {\cal G}_{4\mu} $. 
This is equivalent to
defining ${\cal G}_{4\mu} $ as an off-diagonal operator using
the off-diagonal hopping derivative. Another possibility,
more consistent with Wilson fermions, is a hybrid
hopping derivative that is a combination of the off-diagonal Leibnitz form
and the diagonal form discussed above (see \cite{leib}; this
happens automatically with supersymmetric 
deconstruction in which hopping terms 
are defined as superpotentials).
We will never
need this operator in the derivation of the usual
WZW term in the subsequent section, so these ambiguities
are irrelevant. We will need the fact, however, 
that the diagonal ${\cal G}'_{4\mu}=0$.

Ultimately, such prescriptions codify a number of implicit choices 
of brane configurations and phenomenological outcomes. Unlike the off-diagonal 
antihermitian hopping derivative, the hermitian diagonal one 
preserves topological structures 
associated with chirality (\eg, anomalies). 

\section{Derivation of the WZW Term in the Bianchi Theory}

The CS2 lagrangian may be written in a form more suitable for 
subsequent considerations. Specifically, we start by separating the $A_4$ 
component,
\bea
\label{CStermr}
{\cal{L}}_1 & = & {\cal{L}}_{1a} + {\cal{L}}_{1b}, \cr
{\cal{L}}_{1a} & = & \frac{c}{4}\epsilon^{\mu\nu\rho\sigma}
\Tr(A_4 G_{\mu\nu}G_{\rho\sigma}  ~, 
 + i A_4 A_\mu A_\nu G_{\rho\sigma}
 + i A_4 A_\mu G_{\nu\rho}A_\sigma + i A_A G_{\mu\nu}A_\rho A_\sigma
 \cr
& & \;\;\;\;\; \;\;\;\;\;\;\;\;\;\;
- {2}A_4 A_\mu A_\nu A_\rho A_\sigma )~, 
\cr
{\cal{L}}_{1b} & = &\frac{c}{2}\epsilon^{\mu\nu\rho\sigma}
\Tr(A_\mu G_{\nu\rho}G_{\sigma4} 
 + A_\mu G_{\nu 4 }G_{\rho\sigma }
 + i A_\mu A_\nu A_{\rho}G_{\sigma4} ) .
\eea
This helps re-express ${\cal{L}}_{1a}$ as a lower CS covariant current 
divergence plus an anomaly term,
\bea
\label{CSterm1a}
{\cal{L}}_{1a} & = &  -\frac{c}{2} \Tr(A_4 ~[D_\mu, K^{\mu}]) +
\frac{3c}{4}\epsilon^{\mu\nu\rho\sigma}\Tr(A_4 
G_{\mu\nu}G_{\rho\sigma}), 
\eea
where 
\bea
K^{\mu} \equiv  \epsilon^{\mu\nu\rho\sigma}
\left(iA_\nu A_\rho A_\sigma + G_{\nu\rho}A_\sigma
+ A_\nu G_{\rho\sigma}\right).
 \eea
Likewise, since $G_{\mu 4} = [D_\mu , A_4] -\partial_4 A_\mu$, 
the second term can be written as
\beq
\label{CSterm01b}
{\cal{L}}_{1b} = -\frac{c}{2}\Tr(([D_\mu , A_4] -\partial_4 A_\mu) K^{\mu}). 
\eeq
The combined CS2 is then 
\bea
\label{CS11}
{\cal{L}}_{1} = 
 \frac{c}{2}\Tr ((\partial_4A_\mu)K^\mu)
+\frac{3c}{4}\epsilon^{\mu\nu\rho\sigma}\Tr(A_4 
G_{\mu\nu}G_{\rho\sigma}),
\eea
where some total divergences have been discarded.

Our problem is the interpretation of
the first term above. This problem is obviated when
$G_{\mu 4}=0$, whence we use eq.(\ref{CSterm1a})
for the full lagrangian. We then need to
interpret $[D_\mu, A_4]$.
Consider the definition
of the Wilson line, which we identify with
the chiral field of mesons:
\beq
U = \exp (-i\int A_4 dx^4) = \exp (i\tilde{\pi}) ,
\eeq
where, for a zero-mode $A_4$ we can neglect path-ordering.
We can then write, upon
expanding the $U$'s to second order (this is
the order necessary for for consistent WZW terms---see below):
\beq
\alpha_\mu = U [D_\mu, U^\dagger]
= -i [D_\mu, \tilde{\pi} ] - \half 
(\tilde{\pi} [D_\mu, \tilde{\pi} ] - [D_\mu, \tilde{\pi} ]\tilde{\pi} ) 
+ O(\tilde{\pi}^3), 
\eeq
\beq
\alpha_\mu = U [D_\mu, U^\dagger]
= -i [D_\mu, \int dx^4 A_4] - \half 
(\tilde{\pi} [D_\mu, \int dx^4 A_4] - [D_\mu, \int dx^4 A_4]\tilde{\pi} ) 
+ ...
\eeq
We invert this to make the identification 
\beq
\label{res1}
[D_\mu, \int dx^4  A_4] = i\alpha_\mu 
- \frac{1}{2}
(\tilde{\pi} \alpha_\mu - \alpha_\mu \tilde{\pi} ) 
+ ...  .
\eeq

We now impose the condition that, from our Bianchi-improved theory,
$G_{4\mu} = 0$, equivalently, $\partial_4A_\mu = [D_\mu, A_4]$, and we
substitute eq.(\ref{res1}) into the expression eq.(\ref{CS11}).
The full lagrangian upon integrating over
$x^4$ becomes, 
\bea
\label{CSterm1a}
{\cal{L}}_{1} & = &  i\frac{c}{2}\Tr(\alpha_\mu K^{\mu}) -
\frac{c}{4}\Tr(\tilde{\pi}\alpha_\mu K^{\mu}-\tilde{\pi}K^{\mu}\alpha_\mu ) 
+\frac{3c}{4}\epsilon^{\mu\nu\rho\sigma}\Tr(\tilde{\pi} 
G_{\mu\nu}G_{\rho\sigma}) +... ,
\eea 

We can now check that we recover the Wess-Zumino term.
Turn off the gauge fields, but make the deconstructive
replacement, with the modified vector potential
and field strength summarized in (\ref{trans}), with the primes omitted, 
\beq
A_\mu \rightarrow i\frac{\alpha_\mu}{2},
\qquad \makebox{hence,}
\qquad G_{\mu\nu}\rightarrow -\frac{i}{2}[\alpha_\mu,\alpha_\nu],
\qquad K_\mu \rightarrow
\frac{5}{8}\epsilon_{\mu\nu\rho\sigma}\alpha^\nu\alpha^\rho\alpha^\sigma .
\eeq

Owing to the vector potential which 
is no longer a pure gauge (due to the factor of $1/2$), 
the $G_{\mu\nu}$ terms
are now non-negligible and active in our expression for the 
second Chern character,
and this modifies the WZW term's overall coefficient from
the heuristic argument result in which $G_{\mu\nu}=0$.
The CS term thus becomes on the left end-zone (the $(11)$
matrix element contribution to the trace):
\beq
{\cal{L}}_{1L}  =  
 -\frac{c}{2}\epsilon_{\mu\nu\rho\sigma}\Tr(\tilde{\pi} 
 \alpha^\mu \alpha^\nu \alpha^\rho \alpha^\sigma ) +...  .
\eeq
From the right end-zone, we likewise get the result:
\beq
{\cal{L}}_{1R}  =  
 -\frac{c}{2}\epsilon_{\mu\nu\rho\sigma}\Tr(U^\dagger\tilde{\pi}  U  
 \beta^\mu \beta^\nu \beta^\rho \beta^\sigma )  +... , 
\eeq
which is equivalent, since
$\tilde{\pi} = U\tilde{\pi}U^\dagger$.

Thus, combining, we obtain the Wess-Zumino term
for the Bianchi-consistent theory:
\beq
{\cal{L}}_{1}  =  
 -\frac{N}{240\pi^2}\epsilon_{\mu\nu\rho\sigma}\Tr(\tilde{\pi}
 \alpha^\mu \alpha^\nu \alpha^\rho \alpha^\sigma ) +... ,
\eeq
where the ``index'' $N$ is given by the dimensionality
of the parent theory space-time:
\beq
N = D = 5 . 
\eeq
Evidently, the ``'t Hooft matching'' of our Bianchi
improved theory intrinsically identifies
$D=5$, reflected in the value of
this index. We have no deeper interpretation for
this result at present.

Parenthetically, we may suggest care in manipulating the 
WZ term. For example, we could write (using forms,
and $d\alpha = -\alpha^2$ when the vector
potential is ignored):
\beq
\tr(\tilde{\pi}   \alpha^4) = \tr(\tilde{\pi}   d\alpha d\alpha)
= \tr(d\tilde{\pi}   \alpha^3 ).
\eeq
By naively replacing $d\tilde{\pi}   \rightarrow i\alpha$, we get 
zero for the {\em rhs} by cyclicity of the
$\epsilon$-symbol, $\Tr (\alpha^4) =0$; so we would only access the vanishing, leading 
part of the WZ term: zero! Of course, at the next order in the 
expansion in pions, we recover the properly modified covariant derivative,
\beq
\label{picur}
d\tilde{\pi}    \rightarrow i\alpha
-(1/2)[\tilde{\pi}   , \alpha]   ,
\eeq
and hence consistency for the WZ term, to leading non-trivial order. 

The higher orders for the WZ term have been discussed mathematically in, 
\eg, \cite{BCZ}.  
Beyond leading order, however, the WZ term is not universal in form, as 
an expansion in pions. Indeed, the expansion of unitary chiral fields, such as 
$U=\exp(i\tilde{\pi})$, as a power series
in $\tilde{\pi}$ is non-universal beyond the second
order. (This owes to the fact that pion fields
are ``coordinates'', which parameterize the unitary manifold
satisfying $U^\dagger U =1$. We could equally well have
chosen, \eg, $U= (1+i\tilde{\pi} )/\sqrt{1 + \pi_a\pi^a/f_\pi^2}$. Upon
comparing expansions of both $U$s, it is evident that universality 
is lost at $O(\tilde{\pi}^3)$.) Physically, there is no general way to lock
the coefficients of higher order terms to lower order
ones without additional constraints. Imposing
the equations of motion, however, does lock
the higher order terms to the universal lower order ones
(one must use an expansion in pions in the kinetic term
as well as in the WZ term when the equations of motion
are implemented). The actual on-mass-shell matrix
elements are thus universal.
Consequently, the form of the WZ term is
universal only at the fifth order in $\Tr(\pi\alpha^4)$, since
at the next order we pick up nonuniversal terms from expansions
of $\alpha$. Moreover, there is no way to insure
the self-consistency beyond this order off-mass shell.

\vskip .1in
\noindent
\section{Conclusions}
\vskip .1in

We have initiated the discussion as to how the Chern-Simons terms
of a $D=5 $ pure Yang-Mills theory can be deformed into
the Wess-Zumino-Witten terms of gauged chiral
lagrangians of $D=4$. 

Adjoint currents in $D=5$ are controlled by the second Chern character. 
This in
turn becomes the WZW term in $D=4$. The minimal coefficient of Witten for the
Wess-Zumino term follows from the simplest case of pure gauge
vector potentials generated by London currents in the orbifold 
magnetic superconducting end-zones, as shown in our heuristic argument. 

Singlet currents follow from introduction of a singlet
$U(1)$ vector potential in $D=5$ Yang-Mills,
which is a dual variable describing the
instantonic soliton that uniquely
occurs there. We summarize how this morphs into a new
WZW term in $D=4$, involving the $\sigma$ and $\eta'$ fields,
and which generates the corresponding chiral current equations of
motion. A new $U(1)$ axial current, associated with the $\eta'$, 
has also been identified.
These results are a consequence of the present
approach, and may have application to skyrmion physics.

We then embark upon a formal discussion of the latticization
of the extra (fifth) dimension, and study hopping
derivatives and the Bianchi identities. The coset Bianchi
identity is shown to fail in the case of the diagonal 
hopping derivative in the fifth dimension, 
the most natural definition for a lattice gauge
theory. We find, however, that the coset Bianchi identity can
be rescued if the basic $D=4$ covariant derivative is modified by
the addition of a chiral vector current with the special
coefficient of $1/2$. 

This result has intriguing implications. For one, it converts the
orbifold compactification into an effective periodic compactification. It also
provides a Skyrme term in the effective action that must match the topology of
the instantonic soliton to the skyrmion. We conjecture that with
the fixed coefficient of the Skyrme term provided by the theory,
the matching may be quite powerful, leading perhaps to an analytic skyrmion
solution and some form of ``self-duality.'' 

We finally examine the WZW term implied by the Bianchi-consistent theory. Again
we obtain the WZW term, but now with a coefficient that has an index
of $N=D=5$. Many other issues are raised and future lines to explore 
are suggested by the
present work.

\vskip .2in
\noindent
{\bf Acknowledgments}
\vskip .1in
We thank Nima Arkani-Hamed,
Bill Bardeen, and Lisa Randall for some helpful discussions.
One of us (CTH) thanks the University
of Minnesota's Frontiers II Workshop, and Harvard University
for its hospitality during the final
completion of the present work.
This work is supported in part by
the US Department of Energy, High Energy Physics Division,
Contract W-31-109-ENG-38, and 
grant DE-AC02-76CHO3000.

\end{document}